\documentclass[prc,twocolumn,showpacs,preprintnumbers,nofootinbib,amsmath,amssymb,floatfix,aps]{revtex4-1}
\usepackage{dcolumn}
\usepackage{bm}
\usepackage{color}
\usepackage{amssymb}
\usepackage{amsmath}
\usepackage{graphicx}
\usepackage{amsfonts}
\usepackage{slashed}
\usepackage{pstricks}
\usepackage{float}
\usepackage{hyperref}
\usepackage{array}
\allowdisplaybreaks
\usepackage{dcolumn}
\usepackage{epsf}
\usepackage{natbib}
 \def\openone{\leavevmode\hbox{\small1\kern-3.8pt\normalsize1}}
\begin{document}

\title{Tidal Deformation of Neutron Stars from Microscopic Models of Nuclear Dynamics}

\author{Andrea Sabatucci}
\author{Omar Benhar}
\affiliation{INFN and Dipartimento di Fisica, ``Sapienza'' Universit\`a di Roma, I-00185 Roma, Italy}

\date{\today}

\begin{abstract}
The observation of the gravitational wave signal GW170817, consistent with emission from the inspiral of a binary neutron-star system, 
provided information on the tidal deformation of the participating stars. The available data may be exploited to constrain the equation of state of dense
nuclear matter, as well as to shed light on the underlying models describing nuclear dynamics at microscopic level. 
In this paper we compare the experimental results to the predictions of different theoretical models, based on non relativistic nuclear many-body theory, the relativistic field-theoretical formalism, and a more phenomenological approach constrained by observed nuclear properties. 
While the precision of the available data does not allow to resolve the degeneracy of the models, our analysis shows a distinct 
sensitivity to the star compactness predicted by the different equations of state, which turns out to be significantly affected by relativistic boost
corrections to the nucleon-nucleon potential.
\end{abstract}
%
%\keywords{ Dense matter -- Equation of state -- Stars: neutrons -- Gravitational waves}
{
\index{}\maketitle

%%%%%%%%%%%%%%%%%%%%%%%%
\section{Introduction}
\label{intro}
%%%%%%%%%%%%%%%%%%%%%%%%

On August 17, 2017, the Advanced LIGO-Virgo detector network made the first observation of the gravitational wave signal labelled 
GW170817, consistent with emission from a coalescing binary neutron-star system~\citep{PhysRevLett.119.161101}. The detection 
of this signal, and the later observation of electromagnetic radiation by space- and ground-based telescopes~\citep{Abbott_2017} 
arguably marked the dawning of the long anticipated age of gravitational-wave astronomy.
 
%%%%%%%%%%%%%%%%%%
A great deal of effort is being made to exploit the information extracted from event GW170817 to constrain neutron star (NS) properties, most notably the radius, which are 
in turn related to nuclear matter properties encoded in its Equation of State (EOS), such as the compressibility module and the symmetry energy, see~\cite{baiotti} and references therein. 
The new data will also be critical to the progress of nuclear matter theory, because they provide an unprecedented opportunity to test microscopic models of 
nuclear dynamics in the regime of high density and low temperature, which can not be accessed by terrestrial experiments. 
%%%%%%%%%%%%%%%%%%%

Gravitational waves (GW) emitted during a binary inspiral are driven by the tidal deformation of the participating 
stars, which is largely determined by the nuclear matter EOS~\citep{PhysRevD.81.123016}. However, theoretical studies are often carried out 
using models of the EOS which are only partially derived from a microscopic description of the dynamics of dense nuclear 
matter, see e.g. Ref.~\cite{reddy1}, or simple phenomenological parametrisations based on the information available from measured nuclear properties, see e.g. Ref.~\cite{Bao-AnLi}. 
While the results of these analyses provide valuable information, the extension to the case of  fully microscopic models,  
applicable over the whole range of densities relevant to NSs, is needed to fully exploit the potential of 
GW observations, and shed new light on nuclear dynamics. 
This issue will be all the more important in view of the detection of GW emitted in the aftermath of the excitation of quasi-normal 
modes~\citep{AK1,MNRAS,BFG}, because the interpretation of the signals will require the understanding of NS properties other than 
the EOS, see~\cite{leonardo} and references therein.

In this paper, we analyse the tidal deformation predicted by different neutron star models, to highlight the role played by the description 
of nuclear dynamics at microscopic level. 
The widely employed models that will be referred to as APR1 and APR2 \citep{AP,APR}, as well as the model recently proposed by the authors of Ref.~\cite{Benhar+Lovato.2017}, referred to as BL, have been obtained within the framework of non relativistic Nuclear Many-Body Theory (NMBT), using a nuclear Hamiltonian strongly constrained by the available empirical information on two- and three-nucleon systems. The EOS referred to as GM3, on the other hand, has been derived using the formalism of relativistic quantum field theory and the mean field approximation~\citep{Glendenning.1985,Glendenning+Moszkowski.1991}. This scheme will 
be referred to as Relativistic Mean Field Theory  (RMFT).
For comparison, we have also included in our study a more phenomenological EOS, labelled LS, obtained from extrapolation of nuclear properties
within the conceptual framework of the liquid drop model~\citep{Lattimer+Swesty.1991}. 
The BL, GM3 and LS models have been also recently compared in a study of neutrino luminosity and gravitational wave emission of protoneutron stars~\citep{PhysRevD.96.043015}.

Our study does not include results obtained using the dynamical model based on chiral effective field theory ($\chi$EFT).
While providing an accurate description of the properties of light nuclei, see, e.g., Ref.~\cite{chiral_AFDMC}, chiral potentials
are derived from a low-momentum expansion. They are therefore inherently limited in the ability to describe dense nuclear matter, 
in which nuclear interactions involve large momenta~\cite{benhar:2019}. This problem is highlighted in Ref.~\cite{Tews_ApJ}, whose
authors plainly state that using interactions obtained from $\chi$EFT the EOS of neutron matter can be reliably calculated only
up to one to two times the equilibrium density of isospin-symmetric nuclear matter, $\varrho_0$. In view of the fact that the central density 
of a neutron star of mass $M = 1.4 \ M_\odot$ tipically exceeds $3\varrho_0$, chiral Hamiltonians do not appear to be best suited for 
calculations of neutron star properties.

The main features of the dynamical models of neutron star matter are summarised in Section~\ref{EOS}, while the formalism employed to obtain the tidal deformability is outlined in 
Section~\ref{tidal}. The numerical results of our study are reported and discussed in Section~\ref{results}. Finally, in  Section~\ref{summary} we sum up our findings and state the conclusions.

%%%%%%%%%%%%%%%%%%%%%%%%
\section{Microscopic Models of the Equation of State}
\label{EOS}
%%%%%%%%%%%%%%%%%%%%%%%%

According to NMBT, nuclear matter can be modelled as a collection of pointlike protons and
neutrons, whose dynamics are described by the non relativistic Hamiltonian
\begin{align}
H = \sum_i \frac{p_i^2}{2m} + \sum_{j>i} {v}_{ij} + \sum_{k>j>i} {V}_{ijk}\ ,
\label{hamiltonian}
\end{align}
where $m$ and $p_i$ denote the nucleon mass and momentum, respectively, whereas ${v}_{ij}$ and ${V}_{ijk}$
describe two- and three-nucleon interactions. The two-nucleon potential, that reduces to
Yukawa's one-pion-exchange potential at large distance, is
obtained from an accurate fit to the measured properties of the two-nucleon system, in both 
bound and scattering states, while the purely phenomenological three-body term
$V_{ijk}$ is needed to explain the ground-state energies of the
three-nucleon bound states, and obtain a reasonable account of the empirical equilibrium properties of isospin-symmetric nuclear matter.

The many-body Schr\"odinger equation associated with the Hamiltonian
of Eq.(\ref{hamiltonian}) can be solved exactly, using stochastic Quantum Monte Carlo (QMC) techniques,
for nuclei with mass number $A$ up to 12. The energies
of the ground and low-lying excited states turn out to be in remarkably good agreement with the experimental
data~\citep{QMC}. In the $A \to \infty$ limit, the QMC method has been applied to treat both pure neutron matter (PNM), see Ref.~\cite{QMC}, and, more recently, isospin-symmetric nuclear matter (SNM)~\cite{lonardoni2019}.
Accurate calculations of the ground-state energy can also be performed using the variational method~\cite{AP}. 

In the APR1 model, matter is assumed to consist of neutrons, protons, electrons and muons in $\beta$-equilibrium.
The baryonic equation of state\textemdash constructed  combining PNM and SNM results\textemdash is obtained from a Hamiltonian comprising
 the Argonne ${v}_{18}$ nucleon-nucleon (NN) potential~\citep{WSS} and the 
Urbana IX (UIX) three-nucleon (NNN) potential~\citep{PPCPW}. The expectation value of the Hamiltonian in the ground state, described by 
a trial wave function including correlation effects, is computed using the cluster expansion formalism and chain summation techniques~\citep{AP}.

The APR2 model\textemdash in the literature often referred to as APR\textemdash is similar to the APR1, but takes into account the relativistic correction arising from the boost of the NN potential to a frame 
 in which the total momentum of the interacting pair is non vanishing. These corrections are required to use the 
phenomenological Argonne $v_{18}$ potential\textemdash designed to describe interactions between nucleons in their centre-of-mass frame\textemdash in 
the locally inertial frame associated with the star. 

Inclusion of the boost correction results in the appearance of a sizeable repulsive 
contribution to the potential energy associated with  the NN potential, and to a corresponding reduction of  the repulsion arising from the NNN potential.
The modified NNN potential, to be used in conjunction with the boost-corrected Argonne $v_{18}$ NN potential will be referred to as UIX$^\ast$.

The APR2 EOS of SNM also includes a density-dependent correction to the variational ground-state energy, meant to effectively take into account 
contributions not included in the calculation. This correction, adjusted to reproduce the empirical saturation properties, reaches a maximum 
of 4.5 MeV\textemdash corresponding to $\sim 30$\% of the interaction energy\textemdash at subnuclear density, $\varrho \sim 0.11$~fm$^{-3}$, and rapidly decreases to become negligible in the density region relevant to the NS core.

The impact of the relativistic boost correction on the determination of the potential describing three-nucleon 
forces is illustrated in Fig.~\ref{compare:NNN}. It is apparent that the difference between the potential energy per particle corresponding to the
UIX and UIX$^\ast$ interactions begins to be appreciable just above the equilibrium density of SNM, $\varrho_0 = 0.16$~fm$^{-3}$, and grows 
steeply with $\varrho$.
 
%%%%%%%%%%%%%%%%%%%%%%%%%%%%%%%%%%%%%%%%%%%%%%%%%%%
\begin{figure}[h!]
%\begin{center}
%\vspace*{-.175in}
\includegraphics[width=8.20cm]{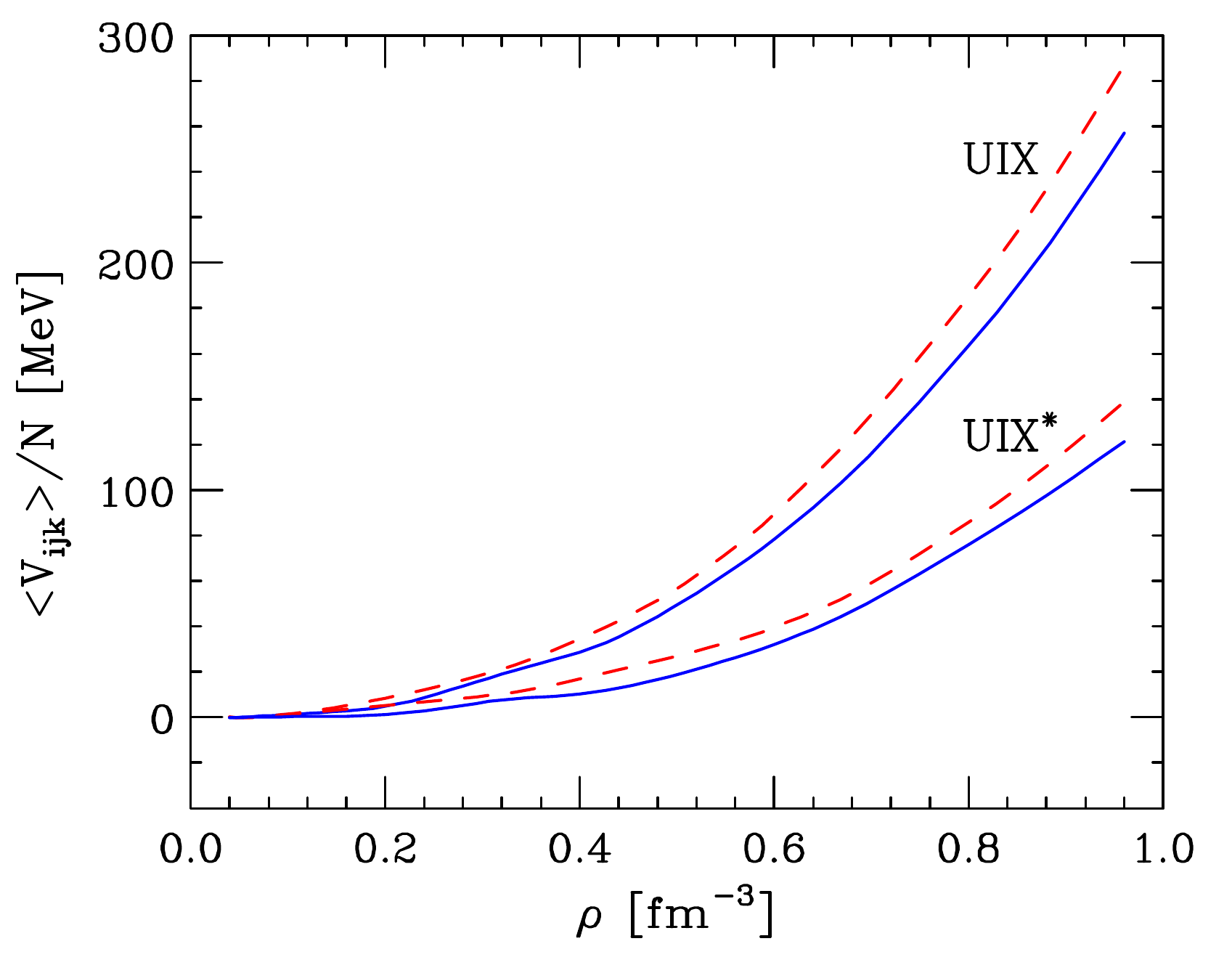}
\vspace*{-.15in}
\caption{Ground-state expectation value of the NNN potential per particle, obtained with (UIX$^*$) and without  (UIX)
inclusion of relativistic boost corrections to the Argonne $v_{18}$ NN potential. 
Solid and dashed lines correspond to  SNM and PNM, respectively.}
\label{compare:NNN}      
% \end{center}
\end{figure}
%%%%%%%%%%%%%%%%%%%%%%%%%%%%%%%%%%%%%%%%%%%%%%%%%%%

The large decrease of the repulsion arising from three-nucleon interactions leads to a softening of the EOS, 
clearly reflected in the density dependence of the pressure of SNM, displayed in Fig~\ref{pressure}.
The solid and dashed lines represent the results obtained from the APR2 and APR1 models, respectively.  
For comparison, the shaded area shows the region consistent with the data obtained from the analysis of nuclear collisions discussed in Ref.~\cite{danielewicz}, providing a constraint on
the pressure  at  $\varrho \geq 2 \varrho_0$. 

%%%%%%%%%%%%%%%%%%%%%%%%%%%%%%%%%%%%%%%%%%%%%%%%%%%
\begin{figure}[h!]
%\begin{center}
%\vspace*{-.175in}
\includegraphics[width=8.25cm]{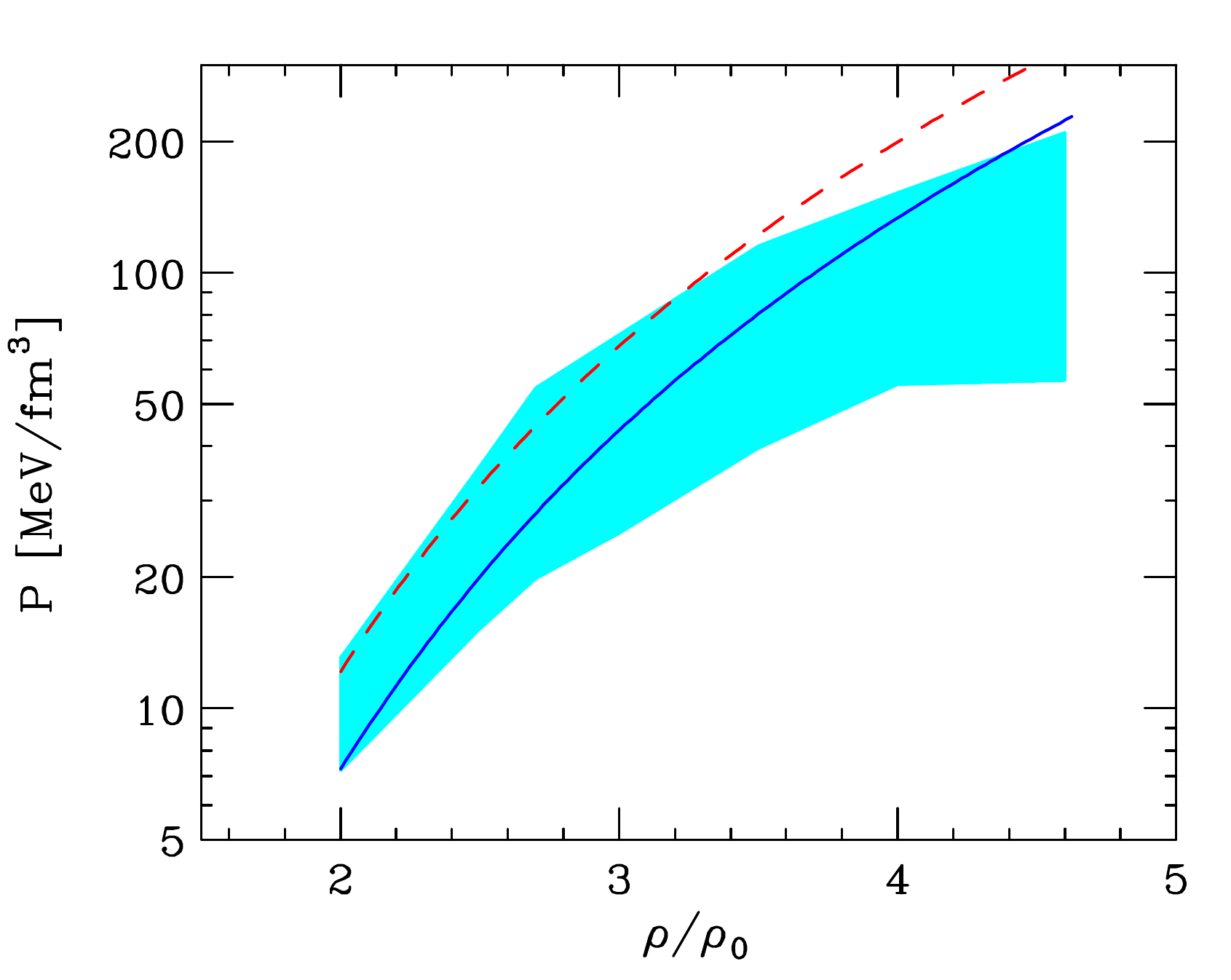}
\vspace*{-.15in}
\caption{Density dependence of the pressure of SNM. The solid line correspond to the APR2 model, including the effect of relativistic boost 
corrections to the NN potential, whereas the dashed line represents the results of the APR1 model. The shaded area corresponds to the
 region consistent with the experimental data reported in Ref.~\cite{danielewicz}. The density is given in units of the equilibrium density
 $\varrho_0 = 0.16$ fm$^{-3}$. }
\label{pressure}      
% \end{center}
\end{figure}
%%%%%%%%%%%%%%%%%%%%%%%%%%%%%%%%%%%%%%%%%%%%%%%%%%%

The BL model exploits the formalism based on correlated wave functions and the cluster expansion to devise 
an {\em effective} NN potential, including the effects of two- and three-nucleon
forces, as well as the density-dependent screening of nuclear interactions arising from strong correlations in coordinate space. 
This effective potential\textemdash  obtained from a bare Hamiltonian comprising the Argonne ${\rm v}_6^\prime$~\citep{AV6P} and UIX potentials\textemdash is well behaved,  and allows to describe the properties of nuclear matter at arbitrary proton fraction 
using standard perturbation theory and the basis of Fermi gas eigenstates \citep{Benhar+Lovato.2017}.

Within RMFT, nucleons are described as Dirac particles interacting through meson exchange.
In the simplest implementation of this scheme the
 dynamics are modelled in terms of the scalar-isoscalar field $\sigma$, that can be identified with a narrow two-pion resonance, 
 and a vector-isoscalar field,  the $\omega$ meson~\citep{QHD1}.
In addition, the GM3 model employed in this study includes the vector-isovector $\rho$ meson~\citep{Glendenning.1985,Glendenning+Moszkowski.1991}.
The equations of motion obtained from this scheme can only 
be solved in the mean field approximation, which amounts to treating  the meson fields
as classical fields. The nuclear matter EOS can then be obtained in closed form, 
and the meson masses and coupling constants appearing in the Lagrangian density
can be determined by fitting the empirical properties of SNM, that is, the 
binding energy, equilibrium density and compressibility.

The models derived within NMBT suffer from the limitations inherent in the non relativistic approximation, 
leading to a violation of causality, determined by the stiffness of the EOS,  in the $\varrho \to \infty$ limit. 
On the other hand, RMFT, while being relativistically consistent 
by construction, is based on a somewhat simplified dynamics, and is not constrained by NN data. Moreover, it 
is plagued by the uncertainty inherent in the use of the mean field
approximation, which is long known to fail in strongly correlated systems~\citep{KB}.

The EOS labelled LS corresponds to the bulk component of the EOS of~\citet{Lattimer+Swesty.1991}.
This model, specifically designed for easy implementation in stellar collapse simulations, has been   
derived from the liquid drop model of the nucleus taking into account the constraints from nuclear 
phenomenology.

Al models considered in our analysis are compatible with the observation of a neutron star of 
mass $M\gtrsim~2 M_\odot$~\citep{demorest}, see Fig.~\ref{fig:1} below.

%%%%%%%%%%%%%%%%%%%%%%%%
\section{Tidal Deformation}
\label{tidal}
%%%%%%%%%%%%%%%%%%%%%%%%

A tide is the deformation of a body produced by the gravitational pull of another nearby body.
Because the deformation depends on the body's internal structure, the observation of tidal effects in binary 
neutron star systems may provide valuable information on the EOS of neutron star matter. 

The orbital motion of two stars gives rise to the emission of gravitational waves (GW), that carry away energy and angular momentum. 
This process leads to a decrease of the orbital radius and, conversely, to an increase of the orbital frequency. 

In the early stage of the inspiral, characterised by large orbital separation and low frequency, the two stars\textemdash of mass $M_1$ and 
$M_2$, with $M_1\geq M_2$\textemdash  behave as point-like bodies and the evolution of the frequency is primarily determined 
by the chirp mass $\mathcal{M}$, defined as \index{Chirp mass}
\begin{equation}\label{teq.1}
\mathcal{M}=\frac{\left(M_1M_2\right)^{3/5}}{\left(M_1+M_2\right)^{1/5}} \ .
\end{equation}

The details of the internal structure become important as the orbital separation approaches the size of the stars. 
The tidal field associated with one of the stars induces a mass-quadrupole moment on the companion, which in turn generates the same 
effect on the first star, thus accelerating coalescence. This effect is quantified by 
the tidal deformability, defined as
\begin{equation}
\label{teq.2}
\Lambda=\frac{2}{3}k_2\left(\frac{c^2R}{GM}\right)^5 \ , 
\end{equation}
where $M$ and $R$ are the star's mass and radius, respectively, and $k_2$ is the second tidal Love number~\citep{0004-637X-677-2-1216}.
For any given stellar mass, the radius and the tidal Love number are uniquely determined by the EOS of neutron star matter.

According to the newtonian theory of gravity, the effect of a quadrupole tidal field is driven by the \emph{tidal momentum}, defined as
\begin{equation}\label{teq.3}
\mathcal{E}_{ij}=-\left.\frac{\partial^2\Phi}{\partial x_i \partial x_j}\right|_{\vec{x}=\vec{r}_c} \ , 
\end{equation} 
where $\Phi$ is the external gravitational potential. The body subject to the tidal momentum, whose centre of mass position is specified by the vector $\vec{r}_c$, develops a quadrupole deformation and the associated 
quadrupole moment 
\begin{equation}\label{teq.4}
Q_{ij}=\int d^3x \;\left(x_ix_j-\frac{1}{3}\delta_{ij}r\right)\varrho(\vec{x}),
\end{equation}
where $\varrho$ is the mass density and $r$ is defined by the equation $r^2=\delta_{ij}x_ix_j$. 

The tensors $Q_{ij}$ and $\mathcal{E}_{ij}$ are both symmetric and traceless. In the  weak field approximation they are related through
\begin{equation}\label{teq.5}
Q_{ij}=-\lambda\,\mathcal{E}_{ij} , 
\end{equation}
and simple dimensional considerations lead to
\begin{equation}\label{teq.6}
\lambda=\frac{2}{3}k_2R^5G^{-1} \ ,
\end{equation}
where the dimensionless constant $k_2$ is the second tidal Love number of Eq.\eqref{teq.2}, and $2/3$ is a conventional factor.

The general relativistic treatment of quadrupole deformations of neutron stars involves the study of linearised perturbations of the equilibrium 
configurations~\citep{1967ApJ...149..591T}. The metric tensor  is written as 
\begin{equation} \label{teq.8}
g_{\alpha\beta}=g_{\alpha \beta}^{(0)}+h_{\alpha \beta} \ ,
\end{equation}
where 
\begin{equation}
g_{\alpha \beta}^{(0)}=\textup{diag}\left(-e^{2\nu(r)},\,e^{2\varphi(r)},\, r^2,\, r^2\sin^2{\theta}\right) \ , 
\end{equation}
is the metric of static and spherically-symmetric spacetime, and the perturbation fulfills the requirement $|h_{\alpha\beta}|<<1$.

Quadrupole effects are associated with the $\ell=2$ even-parity contribution to the expansion of $h_{\alpha\beta}$ in tensorial spherical harmonics, 
whose radial shape is described by the function $H(r)$,  obeying the differential equation~\citep{0004-637X-677-2-1216}
\begin{widetext}
\begin{align}
\label{teq.12}
H''+H'\left\{ \frac{2}{r}+e^{2\varphi}\left[\frac{2M(r)}{r^2}+4\pi r(P-\epsilon)\right] \right\} + 
H\left[-\frac{6e^{2\varphi}}{r^2}+4\pi e^{2\varphi}\left( 5\epsilon+9P+\frac{\epsilon+P}{dP/d\epsilon}\right)-(2\nu')^2\right]=0 \ .
\end{align}
\end{widetext}

Integration of Eq. (\ref{teq.12}) and of the Tolman-Oppenheimer-Volkoff (TOV) equations~\citep{T,OV} allows to determine the 
second tidal Love number, whose expression can be cast in the form
\begin{widetext}
\begin{align}
\label{teq.13}
\nonumber
k_2 =\frac{8}{5} C^5 (1-2C)^2  \big[ 2 + & 2C(y-1) -y \big]  
 \Bigg\{ 2C \big[ 6-3y+3C(5y-8) \big] + 4C^3 \big[13-11y+C(3y-2) \\
& + 2C^2(1+y) \big] +3(1-2C)^2\big[2-y+2C(y-1)\big]\log{(1-2C)}\Bigg\}^{-1} \ , 
\end{align}
\end{widetext}
where $C$ and $y$ are defined as 
\begin{equation}
\label{teq.14}
C=\frac{M}{R},\quad y=R \frac{H'(R)}{H(R)} \ , 
\end{equation}
with $M$ and $R$ being the star mass and radius, respectively.

Equation (\ref{teq.13}) shows that, given a model of the EOS determining the values of $M$ and $R$, 
a calculation of the tidal Love number $k_2$, requires
the knowledge of  the functions $H$ and $H'$, obtained from Eq. (\ref{teq.12}), evaluated at $r=R$. 

%%%%%%%%%%%%%%%%%%%%%%%%%%%%%%%%%%%%%%%%%%%%%%%%%%%
\section{Numerical Results}
\label{results}
%%%%%%%%%%%%%%%%%%%%%%%%%%%%%%%%%%%%%%%%%%%%%%%%%%%

The analysis of the GW170817 signal of \cite{PhysRevLett.119.161101} allowed a precise determination of the chirp mass, the resulting value 
being $\mathcal{M}=1.188^{+0.004}_{-0.002}M_{\odot}$. On the other hand, the estimates of the component masses and their ratio, 
$q=M_2/M_1$, depend on the assumptions made on the NS spins. In this letter, we will consider the results obtained in the 
``low-spin'' scenario, in which the NS spin parameter is restricted to values in agreement with Galactic binary NS measurements.

The mass-radius relations corresponding to the EOSs employed in our study are shown in Fig. \ref{fig:1}. The box
represents the region compatible with the 90\%-confidence-level estimates of mass and radius 
extracted from the analysis of the GW170817 event, yielding $R_1 = R_2 = 11.9 \pm 1.4$ Km, $1.18 \leq M_2 \leq 1.36 \ M_\odot$
and $1.36 \leq M_1 \leq 1.58 \ M_\odot$~\citep{PhysRevLett.121.161101}. 
These values have been obtained using the spectral parametrisation of the EOS~\cite{lindblom}\textemdash constrained to support a NS 
with mass $M \geq 1.97 \ M_\odot$\textemdash at densities $\varrho > \varrho_0/2$, and the Sly EOS of Ref.~\cite{Sly} at lower densities.

%%%%%%%%%%%%%%%%%%%%%%%%%%%%%%%%%%%%%%%%%%%%%%%%%%%
\begin{figure}[h!]
%\begin{center}
%\vspace*{-.175in}
\includegraphics[width=8.0cm]{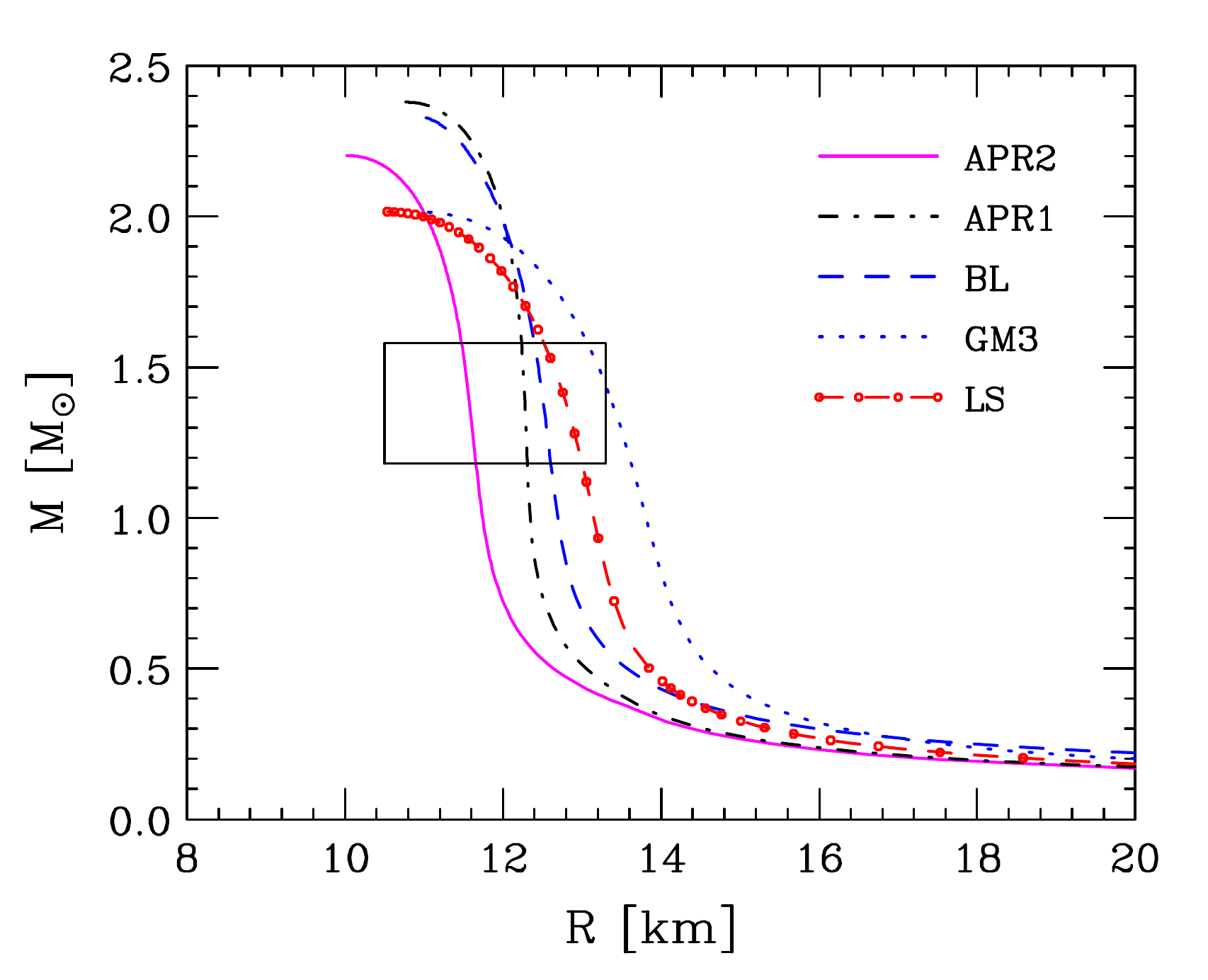}
\vspace*{-.15in}
\caption{Mass-radius relations corresponding to the EOSs employed in this work. The meaning of the labels is explained in Section~\ref{intro}. 
The box represents the 90\%-confidence-level estimate of mass and radius reported by the LIGO-Virgo 
Collaboration~\citep{PhysRevLett.121.161101}. }
\label{fig:1}      
% \end{center}
\end{figure}
%%%%%%%%%%%%%%%%%%%%%%%%%%%%%%%%%%%%%%%%%%%%%%%%%%%

Although the accuracy of the available data does not allow to resolve the degeneracy between the results of different models, it clearly 
appears that the GM3 EOS is only marginally compatible with observations. The curves corresponding to the BL and APR1 
EOSs, obtained from NMBT using similar nuclear Hamiltonians, lie close to one another, while the differences with 
respect to the APR2 model show that relativistic boost correction to the NN potential and the associated modification of the 
NNN potential  result in an appreciable softening of the EOS, see Fig.~\ref{pressure}.

The tidal deformability $\Lambda$\textemdash computed using Eq.~\eqref{teq.2} with 
the values of mass and radius obtained from the EOSs described in Section~\ref{EOS}\textemdash is displayed 
in  Fig.~\ref{fig:2} as a function of the stellar mass. The vertical bar represents the 90\%-confidence-level estimate, 
$70 \leq \Lambda(1.4 \ M_\odot) \leq 580$, obtained by the authors of Ref.~\cite{PhysRevLett.121.161101} by expanding the 
function $M^5 \Lambda(M)$ around $M = 1.4  \ M_\odot$.
The emerging pattern, showing that for any given $M$ the results obtained from different models are ordered according 
to the compactness $C$,  see Eq.\eqref{teq.14}, are consistent with Fig.~\ref{fig:1}. 

%%%%%%%%%%%%%%%%%%%%%%%%%%%%%%%%%%%%%%%%%%%%%%%%%%%
\begin{figure}[h!]
%\begin{center}
\includegraphics[width=8.00cm]{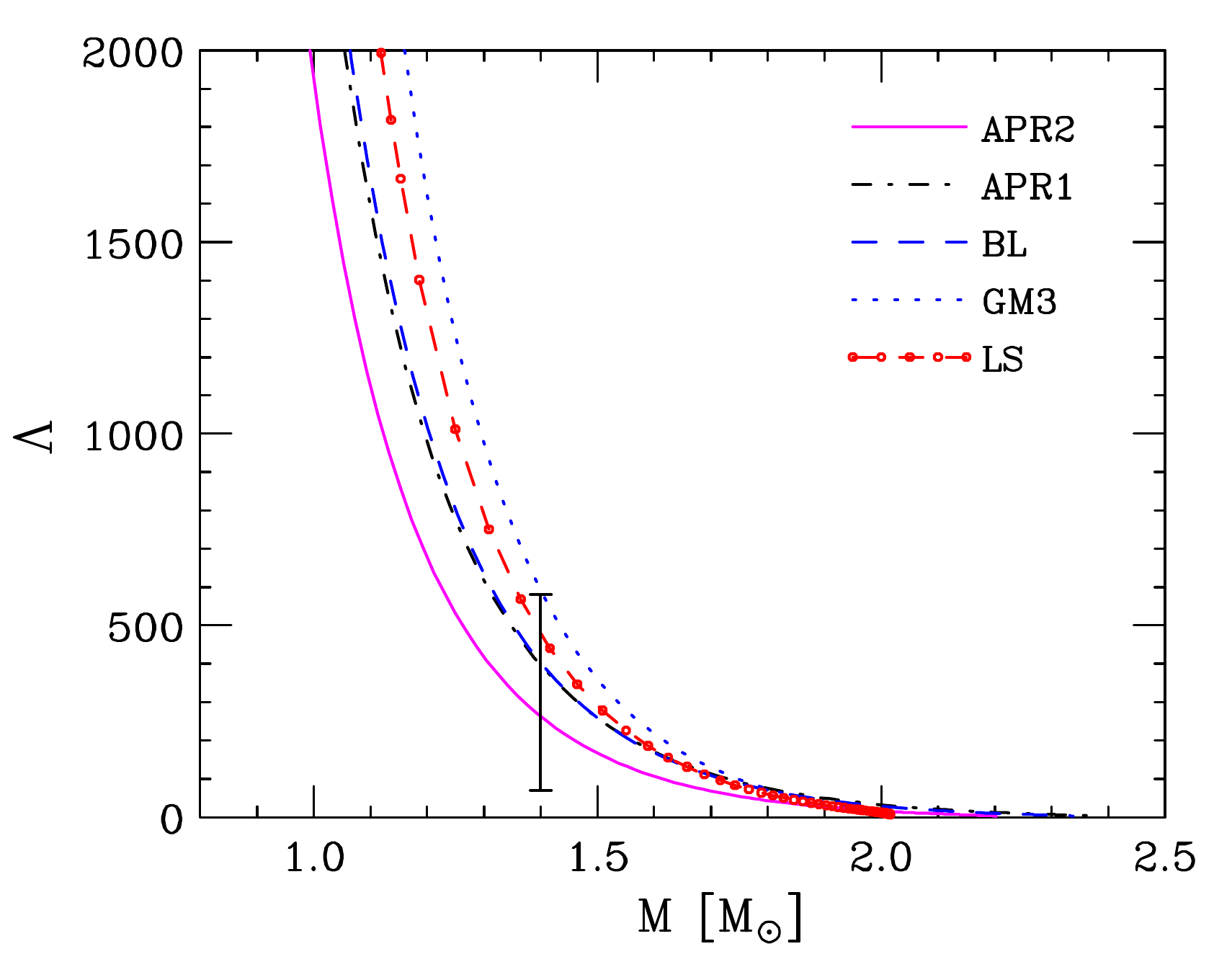}
\vspace*{-.15in}
\caption{ Mass dependence of the tidal deformability obtained from the EOSs 
described in Section~\ref{EOS}, displayed as a function of the stellar mass. 
The vertical bar in the lower panel represents the range of $\Lambda$  for a star of mass $M = 1.4\,M_{\odot}$, extracted from the 
analysis of the GW170817 signal~\citep{PhysRevLett.121.161101}.}
\label{fig:2}      
% \end{center}
\end{figure}
%%%%%%%%%%%%%%%%%%%%%%%%%%%%%%%%%%%%%%%%%%%%%%%%%%%

The authors of Ref.~\cite{PhysRevLett.119.161101} also report the results of an analysis aimed at pinning down the tidal deformability of 
the components of the binary system, $\Lambda_1$ and $\Lambda_2$. 

Assuming a  uniform prior on the quantity  
\begin{equation}
\tilde{\Lambda}=\frac{16}{13}\frac{(M_1+12M_2)M_1^4\Lambda_1+(M_2+12M_1)M_2^4\Lambda_2)}{(M_1+M_2)} \ , 
\end{equation}
determining the GW phase, its value in the low spin scenario has been constrained to $\tilde{\Lambda}\leq 800$ at 90\% confidence level. 
The posterior distribution function for $\Lambda_1$ and $\Lambda_2$ was derived using this constraint, and assuming that both stars in the binary system can be described using the same EOS.

In order to compare theoretical predictions to these data, for each EOS we have generated pairs of stars with masses $M_1$ and $M_2$, distributed 
according to the joint probability distribution reported in Ref.~\cite{PhysRevLett.119.161101} for the low spin scenario.
Because the initial condition for the integration of the TOV equations is the central density $\varrho_0$, not the mass of the star, we have 
solved the equations for a wide range of central densities, to obtain the function $M(\varrho_0)$. Interpolation of this function  
in the region in which $dM/d \varrho_0 \geq 0$, corresponding to stable equilibrium configurations, yields the values of central density of the stars belonging to the binary system, needed to obtain their radii and tidal deformabilities. 

%%%%%%%%%%%%%%%%%%%%%%%%%%%%%%%%%%%%%%%%%%%%%%%%%%%
\begin{table}[h!]
\begin{center}
%\begin{tabular}{p{29mm} p{20mm} p{20mm}}
\begin{tabular}{p{22mm} p{22mm} p{22mm}}
\hline
\hline
EOS &  $ \ \ \ \ R_1 \  [{\rm km}] $ &   $ \ \ \ \ R_2 \  [{\rm km}] $  \\
\hline
\cite{PhysRevLett.121.161101} & $10.50 \  \textendash \ 13.30$ & $10.50 \  \textendash \ 13.30$  \\
\cite{PhysRevLett.123.141101} & $11.98  \  \textendash \ 12.88$ & $11.89 \  \textendash \ 12.98$ \\
APR1 & $ 12.21 \  \textendash \ 12.28 $ 	& $12.28 \  \textendash \ 12.30$  \\

APR2 & $ 11.46 \  \textendash \ 11.58 $  & $ 11.58 \  \textendash \ 11.70 $  \\

BL & $12.38 \  \textendash \ 12.52$ &	$12.52 \  \textendash \ 12.61$    \\

GM3 &		$12.90 \  \textendash \ 13.24$ &		$13.24 \  \textendash \ 13.43$  	\\

LS &		$12.48 \  \textendash \ 12.82$ &		$12.82 \  \textendash \ 13.00$    \\
\hline
\hline
\end{tabular}
\caption{Radii of NSs with masses in the ranges estimated by the authors of Ref.~\cite{PhysRevLett.121.161101}\textemdash $1.36 \leq M_1/M_\odot \leq 1.58$, and $1.18 \leq M_2/M_\odot \leq 1.36$\textemdash computed using the EOSs considered in this work. The first and second row report the radii extracted 
from the analyses of Refs.~\cite{PhysRevLett.121.161101} and \cite{PhysRevLett.123.141101}, respectively.}

\label{3tabR}
\end{center}
\end{table}
%%%%%%%%%%%%%%%%%%%%%%%%%%%%%%%%%%%%%%%%%%%%%%%%%%%

The radii of NSs with masses $M_1$ and $M_2$ within the ranges reported by the authors of Ref.~\cite{PhysRevLett.121.161101}, obtained using the EOSs described in Section~\ref{EOS}, are listed in Table~\ref{3tabR}.

%%%%%%%%%%%%%%%%%%%%%%%%%%%%%%%%%%%%%%%%%%%%%%%%%%%
\begin{figure}[h!]
%\begin{center}
\vspace*{-.10in}
\includegraphics[width=8.80cm]{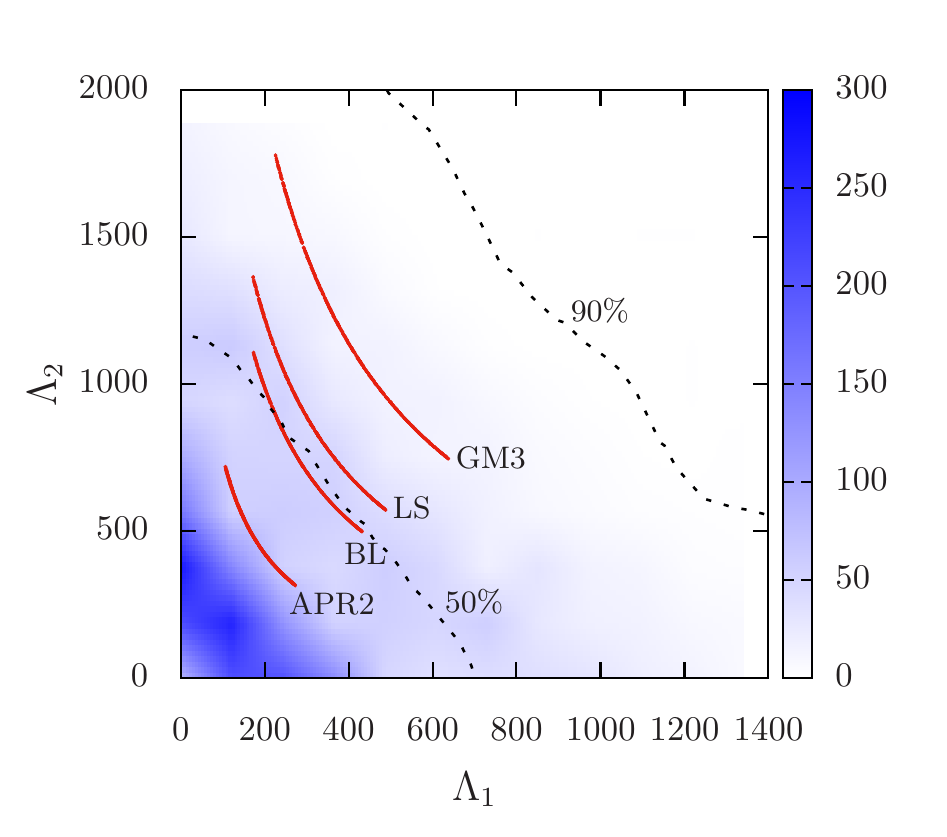}
\vspace*{-.25in}
\caption{Probability density of the tidal deformability parameters, $\Lambda_1$ and  $\Lambda_2$, obtained from the analysis of the GW17081027 signal. The thick solid lines represent the results of calculations carried out using the EOSs described in Section~\ref{EOS}. The dashed lines show  the boundaries of the regions enclosing 50\% and 90\% of the posterior probability density.}
\label{fig:3}      
% \end{center}
\end{figure}
%%%%%%%%%%%%%%%%%%%%%%%%%%%%%%%%%%%%%%%%%%%%%%%%%%%

The results of our calculations of the tidal deformabilities, $\Lambda_1$ and $\Lambda_2$, are displayed in Fig.~\ref{fig:3} together with the data resulting from the analysis performed by the LIGO-Virgo 
Collaboration~\citep{PhysRevLett.119.161101}. The thick curves corresponding to different EOSs are clearly ordered according to compactness, 
and appear to be all compatible with the data at 90\% confidence level. However, only those obtained from NMBT fall within the region 
bounded by the 50\%-confidence-level contour. The curve corresponding to the APR1 model is not included in the figure, because it 
turns out to be nearly indistinguishable from that labelled BL. 

%%%%%%%%%%%%%%%%%%%%%%%%%%%%%%%%%%%%%%%%%%%%%%%%%%%
\section{Summary and Conclusions}
\label{summary}
%%%%%%%%%%%%%%%%%%%%%%%%%%%%%%%%%%%%%%%%%%%%%%%%%%%

We have compared the data obtained from the analysis of the  GW170817 event, detected by the LIGO-Virgo 
Collaboration~\citep{PhysRevLett.119.161101,PhysRevLett.121.161101}, to the predictions of different microscopic  models of nuclear dynamics,  
based on NMBT and RMFT. For the sake of completeness, the results of a more phenomenological approach, derived from the liquid-drop model of the nucleus, have been also included in our study. 

The choice to consider matter consisting of nucleons only appears to be reasonable in view of the masses of the 
coalescing stars, whose values have been estimated to lie in the range $1.2 \ \textendash \ 1.6   \ M_\odot$. In a NS with mass in this range,  
the density is unlikely to  exceed $\sim~3 \varrho_0$\textemdash with $\varrho_0$ being the equilibrium density of isospin-symmetric 
nuclear matter\textemdash and the transition to more exotic phases, involving strange baryons or deconfined quarks, is not expected to occur.

Even though the precision of the available experimental information does not allow to resolve the degeneracy between the predictions of different 
models, our analysis shows a distinct sensitivity to the star compactness, whose value is driven by the EOS and the underlying description of nuclear dynamics. 
Models based on NMBT, in which the dynamics is strongly constrained by the properties of the two-and three-nucleon systems, yield similar predictions, 
as shown by a comparison between the results obtained from the APR1 and BL models. However, the inclusion of relativistic
boost corrections to the  NN potential and the associated modification of the NNN potential result in a softening of the EOS at high density, leading to a sizeable change 
of the mass-radius relation determining the compactness. It is also worth noting that boost corrections push the occurrence of the 
non causal behaviour of the EOS towards higher density, thus expanding the range of applicability of the APR2 model of Ref.~\cite{APR}.
On the other hand, a comparison between the results obtained from NMBT and RMFT suggests that 
the low compactness predicted by the GM3 EOS is likely to be ascribed to the mean-field approximation and to the use of a simplified 
dynamical model, rather than to relativistic corrections to the potential describing NN interactions. 

The possibility to extract more stringent constraints, combining the data collected by the LIGO-Virgo Collaborations with those
obtained from observations of bursts in accreting low-mass x-ray binaries, has been recently investigated 
by the authors of Ref.~\cite{PhysRevLett.123.141101}. While yielding mass ranges close to those reported by~\citep{PhysRevLett.121.161101}, this 
analysis\textemdash based on a phenomenlogical parametrisation of the EOS\textemdash sets more stringent bounds on the radii, $R_1$ and $R_2$, see Table~\ref{3tabR} .

The first observation of GW from a coalescing double NS binary system, and the ensuing developments of the multimessenger approach,
have allowed to obtain valuable new information on the nuclear matter EOS. While being important in their own right\textemdash in that they allow to 
pin down average properties of dense nuclear matter, such as the compressibility and the symmetry energy\textemdash these data 
have the potential to shed light on the underlying dynamics at microscopic level. Future observations with improved sensitivity may  
allow to constrain the NNN potential models in the high-density regime, in which interactions involving more than two nucleons  
become dominant, and shed light on the limits of applicability of the non relativistic approximation.

%%%%%%%%%%%%%%%%%%%%%%%%%%%%%%%%%%%%%%%%%%%%%%%%%%%
\acknowledgments
%%%%%%%%%%%%%%%%%%%%%%%%%%%%%%%%%%%%%%%%%%%%%%%%%%%

The authors are deeply indebted to Andrea Maselli for his many suggestions and continued advice.
Thanks are also due to Valeria Ferrari and Leonardo Gualtieri, for countless illuminating 
discussions on gravitational waves and neutron stars. Finallly, OB gratefully acknowledges the 
hospitality of the Theoretical Physics Department at CERN, where this 
manuscript has been prepared.

%\bibliographystyle{apsrev4-1}
%\bibliography{../tidal}
%merlin.mbs apsrev4-1.bst 2010-07-25 4.21a (PWD, AO, DPC) hacked
%Control: key (0)
%Control: author (72) initials jnrlst
%Control: editor formatted (1) identically to author
%Control: production of article title (-1) disabled
%Control: page (0) single
%Control: year (1) truncated
%Control: production of eprint (0) enabled
%

%
\end{document}